\documentclass[12pt]{article}
\usepackage{axodraw}
\usepackage{cite}

\newcommand{\beq}{\begin{eqnarray}}
\newcommand{\eeq}{\end{eqnarray}}
\newcommand{\drawsquare}[2]{\hbox{%
\rule{#2pt}{#1pt}\hskip-#2pt
\rule{#1pt}{#2pt}\hskip-#1pt
\rule[#1pt]{#1pt}{#2pt}}\rule[#1pt]{#2pt}{#2pt}\hskip-#2pt
\rule{#2pt}{#1pt}}

\newcommand{\Yfund}{\raisebox{-.5pt}{\drawsquare{6.5}{0.4}}}

\newcommand{ \sla }[1]{\setbox0=\hbox{$#1$}         
   \dimen0=\wd0                                     
   \setbox1=\hbox{/} \dimen1=\wd1                   
   \ifdim\dimen0>\dimen1                            
      \rlap{\hbox to \dimen0{\hfil/\hfil}}          
      #1                                            
   \else                                            
      \rlap{\hbox to \dimen1{\hfil$#1$\hfil}}       
      /                                             
   \fi}                                             %

\newcommand{\sfrac}[2]{{\textstyle\frac{#1}{#2}}}

\def\lesssim{\mathrel{\mathpalette\vereq<}}

\def\gtrsim{\mathrel{\mathpalette\vereq>}}
\makeatletter
\def\vereq#1#2{\lower3pt\vbox{\baselineskip1.5pt \lineskip1.5pt
\ialign{$\m@th#1\hfill##\hfil$\crcr#2\crcr\sim\crcr}}}
\makeatother

\def\lesssim{\mathrel{\mathpalette\vereq<}}

\def\gtrsim{\mathrel{\mathpalette\vereq>}}
\makeatletter
\def\vereq#1#2{\lower3pt\vbox{\baselineskip1.5pt \lineskip1.5pt
\ialign{$\m@th#1\hfill##\hfil$\crcr#2\crcr\sim\crcr}}}
\makeatother

\begin{document}

\begin{titlepage}
\begin{center}
    \hfill    MADPH-01-1239\\
    \hfill    HUTP-01/A037\\
    \hfill hep-ph/0107266\\
\vskip .9in
{\LARGE \bf 4D Models of Scherk-Schwarz GUT Breaking} \\
\vspace*{2mm}
{\LARGE \bf via Deconstruction}

\vskip 0.7in
{\bf Csaba Cs\'aki$^{a,}$\footnote{J. Robert Oppenheimer Fellow.},
Graham D. Kribs$^{b}$
 and John Terning$^{c}$}

\vskip 0.15in

$^a${\em Theoretical Division T-8, Los Alamos National Laboratory, Los Alamos,
NM 87545}

\vskip 0.1in

$^b${\em Department of Physics, University of Wisconsin, Madison, WI 53706}

\vskip 0.1in

$^c${\em Department of Physics,
Harvard University, Cambridge, MA 02138}

\vskip 0.1in

{\tt  csaki@lanl.gov, kribs@pheno.physics.wisc.edu, 
terning@schwinger.harvard.edu}

\end{center}

\vskip .5in
\begin{abstract} 
We examine new classes of GUT models where the GUT gauge group is
broken by a 4D analogue of the Scherk-Schwarz mechanism.  These
models are inspired by ``deconstructed'' 5D Scherk-Schwarz
orbifold models. However, no fine tuning of parameters or assumption of 
higher dimensional Lorentz invariance is necessary, and the number 
of lattice sites can be as low as just two.
These models provide simple
ways to solve the doublet-triplet splitting problem, changes proton
decay predictions, and may provide insight into the structure of 
the CKM matrix.
Since the number of fields in these models is finite, the 
corrections to the unification of gauge couplings can be reliably calculated,
and as expected result only in threshold corrections to the differential
running of the couplings. Our analysis  also suggests new
4D models which can enjoy the benefits of orbifold models but
cannot be obtained by deconstruction of a 5D model.
\end{abstract}
\end{titlepage}

\newpage

\section{Introduction}
\setcounter{equation}{0}
\setcounter{footnote}{0}

Unification of the gauge couplings in the supersymmetric extension
of the Standard Model (SM) is one of the strongest experimental
hints both for the existence of supersymmetry (SUSY), and for
grand unified theories (GUTs). However, most SUSY GUT theories 
have several problematic aspects; most notably the doublet-triplet
splitting problem (that is, why are the color triplet Higgs fields
necessary for GUTs are so much heavier than the corresponding doublets),
and the non-observation of proton decay. Even though several possible 
resolutions to the doublet-triplet splitting problem exist in 4D
\cite{missingpartner,otherdt},
the actual implementation of these ideas usually lead to complicated 
models \cite{dtreview}. Recently, new solutions to 
these problems have been suggested by reviving \cite{BHN}
the old Scherk-Schwarz (SS) \cite{SS}
idea of breaking symmetries by boundary conditions in extra dimensional 
models. Many of the basic ideas used in these models
have been discussed within the context of string theory in the 80's
\cite{Witten}. One of the main achievements of the models proposed
recently is to separate the essential features of the Scherk-Schwarz
mechanism from the additional constraints imposed by string theory,
and to try to build minimal realistic models in an effective field theory
context. 
In particular, Kawamura \cite{Kawamura}, Altarelli and
Feruglio \cite{AF} and Hall and Nomura \cite{HN} proposed specific
models with one extra dimension and an $SU(5)$ bulk gauge symmetry. 
(See also \cite{Kobakhidze}. For
earlier work on symmetry breaking by the SS mechanism see \cite{oldSS}.)
In the supersymmetric version of this model, the extra dimension is an
$S^1/Z_2\times Z_2'$ orbifold, where one of the $Z_2$ projections 
reduces the zero mode spectrum to that of 4D ${\cal N}=1$ supersymmetric 
theories, while the second $Z_2$ explicitly breaks the 
GUT symmetry to $SU(3)\times SU(2) \times U(1)$ at the orbifold
fixed-point. 
One of the immediate consequences of this setup is that there is
no need for a field that breaks the GUT symmetry -- the orbifold does that.
Depending on the details of the rest of the setup, most of the 
other problems of SUSY GUTs can also be resolved. In one of the 
simplest implementations proposed by Hebecker and March-Russell \cite{JMR},
all SM fields live at the fixed-point where $SU(5)$ is broken so
there is no need to introduce a Higgs triplet, and proton decay is
absent. Another possibility is to introduce the Higgs in the bulk with
the SM matter fields at one or the other orbifold fixed-points.
In this case the doublet-triplet splitting problem is resolved by the triplets
not having a zero mode, and proton decay is suppressed arranging that 
dangerous dimension five operators are absent due to the structure of 
the triplet masses enforced by global symmetries \cite{HN}. 
For more recent work on SS breaking of symmetries see 
\cite{differential,finiteHiggs,regularization,BHN2,Faraggi,orbifoldbreaking,
Haba}.

Recently it has been realized, that 
many seemingly extra dimensional ideas can be implemented 
within a purely 4D theory by considering a ``deconstructed'' 
(or latticized) version of the extra dimension
\cite{ACG,Fermi1}. This construction can be used to obtain 
a variety of interesting 4D models \cite{deconstr}. 
The aim of this paper is to give simple 4D 
models that mimic the above outlined SS-type breaking of the GUT
symmetry. For this we will use the supersymmetric versions of deconstructed
extra dimensions obtained in \cite{CEGK} based on the earlier work
of \cite{CEFS}. We will present several different GUT models. In all of them 
there will be $N-1$ copies (where $N$ can be as low as just two)
of the $SU(5)$ gauge group, and a single
$SU(3)\times SU(2)\times U(1)$ group, which will be broken to the
diagonal SM gauge group. The explicit breaking of the $SU(5)$ symmetry
at the last link is the deconstructed version of the orbifold point
breaking the GUT symmetry. Using this basic setup we will obtain 
different models depending on how the Higgs and SM matter fields are
introduced. First we will include both the Higgs and the matter fields only
into the last SM gauge group, while later we will present various 
modifications of the simplest model. Since there are a finite number of states
in this theory, the correction to the unification of gauge couplings can be
reliably estimated.  This correction to the difference of the gauge
couplings is of the same form as the threshold correction
found in the continuum case.  Note that after this work
was completed, we learned that similar ideas have been pursued
by Cheng, Matchev, and Wang \cite{ChengWang}.

\section{GUTs without Higgs Triplets}
\setcounter{equation}{0}
\setcounter{footnote}{0}

\subsection{The simplest model}

The first model that we will consider is the simplest possible 
construction based on the higher dimensional example presented
in \cite{JMR}.  In this 5D model the bulk has eight supercharges 
(${\cal N}=2$ supersymmetry in 4D). Half of these supersymmetries
are broken by the first $Z_2$ orbifold projection, while the second
projection breaks $SU(5)$ to the SM gauge group. All MSSM matter and
Higgs fields are included on the second orbifold fixed-point where the
gauge groups is reduced to the SM group.  

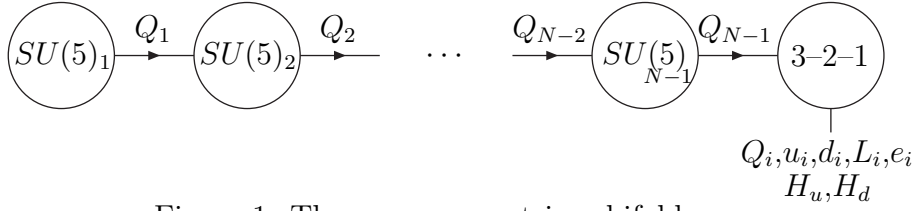
\begin{figure}[t]
\begin{picture}(400,60)(0,0)
  \CArc(  80, 40 )( 20, 0, 360 )
  \Text(  81, 40 )[c]{$SU(5)_1$}
  \ArrowLine( 100, 40 )( 130, 40 )
  \Text( 115, 50 )[c]{$Q_1$}
  \CArc( 150, 40 )( 20, 0, 360 )
  \Text( 151, 40 )[c]{$SU(5)_2$}
  \ArrowLine( 170, 40 )( 200, 40 )
  \Text( 185, 50 )[c]{$Q_2$}
  \Text( 225, 40 )[c]{$\cdots$}
  \ArrowLine( 250, 40 )( 280, 40 )
  \Text( 265, 50 )[c]{$Q_{N-2}$}
  \CArc( 300, 40 )( 20, 0, 360 )
  \Text( 301, 40 )[c]{$SU(5)$}
  \Text( 310, 32 )[c]{$_{N-1}$}
  \ArrowLine( 320, 40 )( 350, 40 )
  \Text( 335, 50 )[c]{$Q_{N-1}$}
  \CArc( 370, 40 )( 20, 0, 360 )
  \Text( 371, 40 )[c]{3--2--1}
  \Line( 370, 20 )( 370, 11 )
  \Text( 370,  3 )[c]{$Q_i$,$u_i$,$d_i$,$L_i$,$e_i$}
  \Text( 370, -10 )[c]{$H_u$,$H_d$}
\end{picture}
\caption{The supersymmetric orbifold moose.}
\label{N-cite-fig}
\end{figure}

A simple construction for the $S^1/Z_2$ orbifold with 4 supercharges 
in 4D has been presented in \cite{CEGK}. The construction is
summarized below:

\begin{equation}
\label{orbifold}
\begin{array}{c|ccccc}
      & SU(M)_1 & SU(M)_2 & \cdots & SU(M)_{N-1} & SU(M)_N \\
\hline
{\vrule height 15pt depth 5pt width 0pt}
 Q_1       & \Yfund            & \overline{\Yfund}  & 1       & \cdots & 1 \\
  Q_2       & 1       & \Yfund  & \cdots & 1 & 1 \\
  \vdots    & \vdots & \vdots & \ddots & \vdots & \vdots \\
  Q_{N-1}   & 1 & 1 & \cdots & \Yfund & \overline{\Yfund} \\
\bar{P}_{1,\ldots ,M} &
\overline{\Yfund} & 1 & \cdots & 1 & 1 \\
  P_{1,\ldots ,M} & 1 & 1 & \cdots & 1 & \Yfund\\
\end{array}. \nonumber
\end{equation}
The bifundamentals $Q_i$'s obtain a vacuum expectation value (VEV) 
either due to some 
strong interactions or due to a superpotential that 
breaks the gauge group to the diagonal $SU(M)$. The fields
$P_i$ and $\bar{P}_i$ are necessary only for anomaly cancellation
in the endpoint gauge groups, obtaining a mass from the superpotential term
\begin{equation}
\frac{1}{M_{\rm Pl}^{N-2}}
\sum_{i=1}^{M} \bar{P}_i \prod_{j=1}^{N-1} Q_j P_i.
\end{equation}
In order to obtain the Scherk-Schwarz-type GUT breaking in this model,
we take all $SU(M)$'s to be given by $SU(5)$ (this
can be easily generalized to larger GUT groups), except the last gauge
group is replaced by the SM gauge group $SU(3)\times SU(2)\times U(1)$.
We illustrate this model in Fig.~\ref{N-cite-fig}.
This replacement is supposed to mimic the effect of the second orbifold
projection that explicitly breaks the GUT symmetry. As explained above, 
in this simplest version the SM matter fields are included at the orbifold
fixed-point, thus in our case they will only transform under the last 
gauge group. Therefore the matter content of the theory is thus given by

\begin{equation}
\label{simpless}
\begin{array}{c|ccccccr}
      & SU(5)_1 & SU(5)_2 & \cdots & SU(5)_{N-1} & SU(3) & SU(2) & U(1)_Y  \\
\hline
{\vrule height 15pt depth 5pt width 0pt}
 \bar{P}_{1,\ldots ,5} &
\overline{\Yfund} & 1 & \cdots & 1 & 1 & 1 & 0 \\
Q_1  & \Yfund  & \overline{\Yfund}  & 1       & \cdots & 1 & 1 & 0\\
Q_2  & 1       & \Yfund  & \cdots & 1 & 1 & 1 & 0 \\
        \vdots & \vdots & \vdots & \ddots & \vdots & \vdots & \vdots& \vdots \\
Q_{t,N-1}   & 1 & 1 & \cdots & \Yfund & \overline{\Yfund} & 1 &\frac{1}{3}\\
Q_{d,N-1}   & 1 & 1 & \cdots & \Yfund & 1 & \Yfund & -\frac{1}{2}\\
P_{1,\ldots ,5} & 1 & 1 & \cdots & 1 & \Yfund & 1 & -\frac{1}{3}\\
P'_{1,\ldots ,5}& 1 & 1 & \cdots & 1 & 1 & \Yfund & \frac{1}{2}\\
H_u & 1 & 1 & \cdots & 1 & 1 & \Yfund & \frac{1}{2} \\
H_d & 1 & 1 & \cdots & 1 & 1 & \Yfund & -\frac{1}{2} \\
L_i & 1 & 1 & \cdots & 1 & 1 & \Yfund & -\frac{1}{2} \\
E_i & 1 & 1 & \cdots & 1 & 1 & 1 & 1 \\
Q_i & 1 & 1 & \cdots & 1 & \Yfund & \Yfund & \frac{1}{6} \\
\bar{U}_i & 1 & 1 & \cdots & 1 & \overline{\Yfund} & 1 & -\frac{2}{3} \\
\bar{D}_i & 1 & 1 & \cdots & 1 & \overline{\Yfund} & 1 & \frac{1}{3} \\
\end{array}. \nonumber
\end{equation}

Carefully inspecting this model one immediately recognizes, that for
$N=2$ this theory is identical to the model proposed recently by
Weiner \cite{Weiner}, who purely from 4D considerations wrote down this 
model as an example of a theory that allows the unification of gauge 
couplings without actually unifying the representations. Thus Weiner's model
is the simplest 4D realization of the Scherk-Schwarz 
breaking of the GUT symmetries. Therefore, all the features
that Weiner's model possesses apply in the general $N$ case here as
well. The doublet-triplet splitting is resolved by not embedding
the SM fields (in particular, the Higgs doublets) into 
into unified multiplets.
Proton decay is absent, since the SM fields do not interact with the 
$X,Y$ gauge bosons that transform as $(3,2)$ under $SU(3)\times SU(2)$.

One could include an $SU(5)$ on the last site instead of  
$SU(3)\times SU(2)\times U(1)$ if an $SU(5)$ adjoint field $\Sigma$
was added such that it obtains a large ($\gg M_{GUT}$) VEV
that breaks the $SU(5)$ to $SU(3)\times SU(2)\times U(1)$. This would then
be the deconstructed version of the model presented in \cite{differential}. 
The advantage of this approach is that charge quantization is explained, 
but then the doublet-triplet splitting has to be explained with one of the
conventional methods. Proton decay will be still very much suppressed due to
the large value of the $\Sigma$ VEV. Unification of couplings will still not 
be exact, due to the fact that operators involving the $\Sigma$ field will
give order one $SU(5)$ non-symmetric corrections to the gauge couplings
at the high scale. We will not consider this possibility any further in
this paper.

The fields $P,\bar{P}$ needed to cancel the endpoint gauge anomalies
must obtain a mass.  In the simplest version of the model, non-renormalizable
operators of the form $\frac{1}{M_{Pl}^{N-2}} \bar{P} Q_1\ldots Q_{N-1} P$,
result in mass terms of the order $v (\frac{v}{M_{Pl}})^{N-2}$,
with $v$ of order the GUT scale.  The $P,\bar{P}$ fields 
will appear as massive doublets and triplets in the theory below the 
unification scale. Since they form complete $SU(5)$ multiplets, they 
do not affect the unification of couplings, as long as their masses
are reasonably large. Requiring that these masses are above the TeV scale
will restrict the number of lattice sites to $N \lesssim 9$. However, this 
constraint is not very robust. If the $Q$ fields themselves are composites,
then one gets a stronger suppression of the masses and a stronger bound on
$N$. However, since the $P,\bar{P}$ fields are not an essential part
of the construction, they are added only for anomaly cancellation, one 
could modify the model such that there are no $P,\bar{P}$ fields at all,
at the price of for example doubling the number of link fields in the 
theory.

\subsection{Gauge coupling unification}
\label{GCU-sec}

Gauge coupling unification is a success in the MSSM, so it is
natural to ask whether gauge coupling unification is maintained
in this model, and at what scale the couplings are unified.
Here we take the VEVs of all the bifundamental fields $Q_i$ to be 
equal to $v$, and all the gauge couplings of the $SU(5)$ groups 
to be equal to $g$.  This implies the ``spacing'' between lattice 
sites $a^{-1} = \sqrt{2} g v$ is the same across the 
entire lattice.  We do this to simplify the calculations
of the mass matrices.  Our construction does {\em not}\/ require 
a ``hopping'' symmetry nor any other remnant of higher dimensional 
Lorentz invariance.  In the MSSM, unification 
occurs roughly at the mass scale of the physical $X,Y$ gauge bosons.
Here, unification occurs at the scale where the diagonal subgroup
is no longer a physical description of the model.  This occurs 
once the lattice is resolved, meaning the scale of heaviest 
KK excitation, which is $2 a^{-1}$.  Once above this scale, the KK modes 
become the dynamical fields associated with the full product gauge theory.
Thus, it is at this scale that one can perform a matching of the gauge 
couplings between the diagonal subgroup and the $N$ lattice sites,
\begin{eqnarray}
\label{matching}
&& \frac{1}{g_{SU(3),{\rm diag}}^2}= \sum_{i=1}^{N-1} \frac{1}{g_i^2} +
\frac{1}{g_{SU(3)}^2}, \nonumber \\ 
&& \frac{1}{g_{SU(2),{\rm diag}}^2}= \sum_{i=1}^{N-1} \frac{1}{g_i^2} +
\frac{1}{g_{SU(2)}^2}, \nonumber \\ 
&& \frac{1}{g_{U(1),{\rm diag}}^2}= \sum_{i=1}^{N-1} \frac{1}{g_i^2} +
\frac{1}{g_{U(1)}^2} \; .
\end{eqnarray}
Thus we can see that the same quantity $\sum_{i=1}^{N-1} \frac{1}{g_i^2}$
arises in all three diagonal subgroup gauge couplings.  It is only
through the explicit breaking of the gauge symmetry at the last site 
that one obtains a contribution which is generically not universal, 
and so the unification of gauge couplings in these models is not exact. 
These non-universal terms are the exact analogs of
the possible brane localized kinetic terms that can be added in the
continuum theory which can alter unification.  However,
if the number of gauge groups is significant {\em and}\/ 
the last site couplings are not too different from the other $SU(5)$
couplings, then one expects the diagonal subgroup couplings 
are approximately unified.  Thus for a large number of $SU(5)$ 
gauge groups one does not 
need to ensure that the $SU(3)\times SU(2)\times U(1)$ on
the last site is much more strongly coupled than the $SU(5)$'s, as is the
case of $N=2$. This is the analog of $SU(5)$ breaking brane localized 
terms in the continuum theory that are volume suppressed compared 
to the $SU(5)$ symmetric terms.

Below the scale $2 a^{-1}$, the gauge group resulting from the 
diagonal breaking of $[SU(5)]^{N-1} \times$ SM group is 
just the SM.  Between $2 a^{-1}$ and $a^{-1}/N$, the theory appears five
dimensional with a KK-like tower of nearly $SU(5)$ symmetric states.
(Hereafter, we simply use ``KK tower'' to refer to the gauge
and gaugino excitations resulting from the diagonal breaking of the
gauge symmetries forming the lattice.)  The $N \times N$ gauge boson mass
matrix for the SM is
\begin{equation}
4 a^{-2} \left( \begin{array}{rrrrr} 
        1 & -1 &        &    &    \\
       -1 &  2 &   -1   &    &    \\ 
          &    & \ddots &    &    \\ 
          &    &   -1   &  2 & -1 \\ 
          &    &        & -1 &  1 \\ 
       \end{array} \right)
\end{equation}
with eigenvalues 
\begin{eqnarray}
m_{3,2,1} &=& \frac{2}{a} \sin \frac{j \pi}{2 N} \; ,
\end{eqnarray}
where $j=0,1,2 \ldots , N-1$.
The gaugino and scalar adjoint masses are identical to the gauge 
bosons for $j>0$ \cite{CEGK}.  The $(N-1) \times (N-1)$ mass matrix of the 
$X,Y$ gauge bosons (and associated fermions and scalars) is \cite{CEGKT}
\begin{equation}
4 a^{-2} \left( \begin{array}{rrrrr} 
        1 & -1 &        &    &    \\
       -1 &  2 &   -1   &    &    \\ 
          &    & \ddots &    &    \\ 
          &    &   -1   &  2 & -1 \\ 
          &    &        & -1 &  2 \\ 
       \end{array} \right)
\end{equation}
with eigenvalues
\begin{eqnarray}
m_{X,Y} &=& \frac{2}{a} \sin \frac{(2 k + 1) \pi}{4 N - 2} \; ,
\end{eqnarray}
where $k=0,1, \ldots , N-2$.
These masses are the latticized analogs of the KK mass spectrum
found in Ref.~\cite{HN}.  This can be seen by writing the approximate
masses of the low lying modes, assuming $N$ is large, as 
\begin{eqnarray}
m_{3,2,1} \sim \frac{2 n + 2}{R} \\
m_{X,Y} \sim \frac{2 n + 1}{R} 
\end{eqnarray}
with the identification $R = 2 a N/\pi$.

We can estimate the evolution of the diagonal subgroup gauge couplings
below the strong coupling scale via one-loop renormalization
group equations in a manner analogous to Refs.~\cite{HN,JMR}.  
The low energy couplings are
\begin{equation}
\alpha_i^{-1}(M_Z) = \alpha_i^{-1}(2 a^{-1}) + \frac{1}{2\pi} 
    \left[   b_i \ln \frac{2 a^{-1}}{M_Z}
           + c_i \sum_{n=1}^{N-1} \ln \frac{2 a^{-1}}{2 a^{-1} 
             \sin \frac{n \pi}{2 N}}
           + d_i \sum_{n=0}^{N-2} \ln \frac{2 a^{-1}}{2 a^{-1} 
             \sin \frac{(2 n + 1) \pi}{4 N - 2}} 
    \right] \label{evolve-one-eq}
\end{equation} 
where $b_i = (33/5,1,-3)$, $c_i = (0,-4,-6)$, $d_i = (-10,-6,-4)$
are the beta-function coefficients for the MSSM, one SM KK excitation,
and one X,Y KK excitation.  The split $SU(5)$ multiplets are of course
a direct consequence of breaking $SU(5)$ on one lattice site.

We can convert the sums over the KK excitation masses into products
of sines as
\begin{eqnarray}
\sum_{n=1}^{N-1} \ln \frac{a^{-1}}{a^{-1} \sin \frac{n \pi}{2 N}}
 &=& - \frac{1}{2} \ln \, \prod_{n=1}^{N-1} \sin^2 \frac{n \pi}{2 N} \\
\sum_{n=0}^{N-2} \ln \frac{a^{-1}}{a^{-1} \sin \frac{(2 n + 1) \pi}{4 N - 2}}
 &=& - \frac{1}{2} \ln \, \prod_{n=0}^{N-2} \sin^2 \frac{(2 n + 1) \pi}{4 N -2}
\end{eqnarray}
The products of sines take the simple forms
\begin{eqnarray}
\prod_{n=1}^{N-1} \sin^2 \frac{n \pi}{2 N} &=& \frac{4 N}{2^{2 N}} \\
\prod_{n=0}^{N-2} \sin^2 \frac{(2 n + 1) \pi}{4 N - 2} &=& \frac{4}{2^{2 N}}
\end{eqnarray}
which we can use to rewrite the evolution equation (\ref{evolve-one-eq}) 
as 
\begin{equation}
\label{coupling}
\alpha_i^{-1}(M_Z) = \alpha_i^{-1}(2 a^{-1}) + \frac{1}{2\pi} 
    \left[   b_i \ln \frac{2 a^{-1}}{M_Z}
           + (c_i + d_i) (N - 1) \ln 2 
           - \frac{c_i}{2} \ln N
    \right] \; .
\end{equation} 
The sum $c_i + d_i = (-10,-10,-10)$ is independent of $i$, 
and thus the second term represents the $SU(5)$ {\em symmetric}\/ power 
law running \cite{DDG} between $a^{-1}/N$ and $2 a^{-1}$.  The last term 
proportional to $\ln N$ is different for each gauge coupling,
representing the logarithmic differential running \cite{differential}
between $a^{-1}/N$ and $2 a^{-1}$.  Thus one can see from (\ref{coupling})
that 
each gauge coupling receives a different logarithmic 
dependence on the number of lattice sites.  Since $N$ sets the 
ratio of the unification scale to the scale where the first KK 
excitations appears, it is clear that this effect is entirely
analogous to an ordinary 4D GUT scale threshold correction
resulting from GUT fields with masses below the unification scale.

As we have discussed, unification (or at least approximate
unification) is expected to occur at $2 a^{-1}$.  We can 
calculate this scale for a given number $N$ of lattice sites
by equating $\alpha_i(2 a^{-1}) = \alpha_j(2 a^{-1})$
for fixed $i,j$ and solving for the unification scale.  Using the 
one-loop evolution equations from above, we find
\begin{equation}
\alpha_i^{-1}(M_Z) - \alpha_j^{-1}(M_Z) = \frac{1}{2\pi} 
    \left[   (b_i - b_j) \ln \frac{2 a^{-1}}{M_Z}
           - \frac{c_i - c_j}{2} \ln N
    \right] \; . \label{diff-eq}
\end{equation}
Of course any two gauge couplings intersect, leading to
the one-loop MSSM result
\begin{equation}
\alpha_i^{-1}(M_Z) - \alpha_j^{-1}(M_Z) = \frac{1}{2\pi} 
    (b_i - b_j) \ln \frac{M_{ij}}{M_Z} \; .
\end{equation}
where $M_{ij}$ is the intersection scale.  
We can therefore rewrite (\ref{diff-eq}) as
\begin{equation}
0 = \frac{1}{2\pi} 
    \left[   (b_i - b_j) \ln \frac{2 a^{-1}}{M_{ij}}
           - \frac{c_i - c_j}{2} \ln N
    \right] \; ,
\end{equation}
leading to a relationship between $2 a^{-1}$ and $M_{ij}$,
\begin{equation}
2 a^{-1} = N^{\frac{c_i - c_j}{2 (b_i - b_j)}} M_{ij}
\qquad \mathrm{with} \qquad 
\frac{c_i - c_j}{2 (b_i - b_j)} = \left\{ 
   \begin{array}{ll}
   5/14 & \quad \{ij\} = \{12\} \\
   5/16 & \quad \{ij\} = \{13\} \\
   1/4  & \quad \{ij\} = \{23\}
   \end{array}
\right .
\end{equation}
If the MSSM gauge coupling unification were exact, meaning 
$M_{ij} = M_{\rm GUT}$, the gauge couplings here would never 
unify exactly for $N > 1$.  Of course the MSSM gauge coupling
unification is only accurate to a few percent.  In fact,
unification can be more precise in this model than in the MSSM
since, as we will see, the correction to $\alpha_3$ is negative.
One reasonable approach is to require $\alpha_1 = \alpha_2$ at the 
unification scale, since they are the most accurately measured couplings 
at the weak scale.  Defining $M_{12} = M_{\rm GUT}$, this implies
\begin{eqnarray}
2 a^{-1} &=& N^{5/14} M_{\rm GUT} \label{last-KK-mass-eq} \\ 
\frac{a^{-1}}{N} &=& \frac{1}{2 N^{9/14}} M_{\rm GUT} \; .
    \label{first-KK-mass-eq}
\end{eqnarray}
Thus, the scale where the KK-like states begin to appear is
{\em lower}\/ than the usual MSSM unification scale by a 
factor $1/(2 N^{9/14})$, while the scale of 
gauge coupling unification is {\em higher}\/ by a factor
$N^{5/14}$.  We can then calculate the deviation of $\alpha_3$
from $\alpha_1 = \alpha_2 = \alpha_{\rm GUT}$,
\begin{eqnarray}
\frac{\alpha_{\rm GUT} - \alpha_3}{\alpha_{\rm GUT}} \left|_{2 a^{-1}}\right . 
    &\simeq& \frac{3}{14\pi} \alpha_{\rm GUT} \, \ln N \; .
\end{eqnarray}
This is a rather small correction.  For example, $\alpha_3$
is unified with the other gauge couplings 
to $< 1\%$ accuracy for $N < 40$.  However, amusingly the sign
of this correction is such that $\alpha_3(M_Z)$ is closer to the
experimentally measured value and so unification in this model
may be more precise than in the MSSM for reasonably small values of $N$.

Note that the value of the unified coupling $\alpha_{\rm GUT}$ is 
smaller than the usual GUT coupling $\alpha_{\rm GUT} \sim 1/25$ 
as is expected from adding vector supermultiplets to the diagonal subgroup.
Of course the 't Hooft coupling $\alpha N$ is increasing for 
larger $N$.  Very roughly the diagonal subgroup becomes strongly
coupled once $\alpha N \sim 1$, which suggests $N$ cannot be much larger 
than about $25$ to maintain a perturbative analysis.

\subsection{An example of dynamical breaking to the diagonal group}

Let us now give a concrete realization of the strong dynamics that can
enforce the expectation values for the bifundamentals $Q_i$. We 
discuss only the $N=2$ case in detail; the cases with larger $N$ can be 
trivially generalized from this. We consider an $SU(5)$ SUSY gauge theory
with five flavors, and assume that that the above $SU(5)\times SU(3)\times
SU(2)\times U(1)$ is the weakly gauged subgroup of the global symmetries.
Thus the full matter content is given by
\begin{equation}
\label{N=2}
\begin{array}{c|ccccc}
      & SU(5) & SU(5)& SU(3) & SU(2) & U(1) \\
\hline
{\vrule height 15pt depth 5pt width 0pt}
q & \Yfund & \overline{\Yfund} & 1 & 1 & 0 \\
\bar{q} & \overline{\Yfund} & 1 & \Yfund & 1 & - \frac{1}{3}\\
\bar{q}'& \overline{\Yfund} & 1 & 1 & \Yfund & \frac{1}{2}\\
\bar{P}_1, \dots, \bar{P}_5 & 1 & \Yfund & 1 & 1 & 0 \\
P_1, \dots, P_5 & 1 & 1 & \overline{\Yfund} & 1 & \frac{1}{3}\\
P_1', \dots, P_5' & 1 & 1 & 1 & \Yfund & -\frac{1}{2}\\
H_u & 1 & 1 & 1 & \Yfund & \frac{1}{2} \\
H_d & 1 & 1 & 1 & \Yfund & -\frac{1}{2} \\
L_i & 1 & 1 & 1 & \Yfund & -\frac{1}{2} \\
E_i & 1 & 1 & 1 & 1 & 1 \\
Q_i & 1 & 1 & \Yfund & \Yfund & \frac{1}{6} \\
\bar{U}_i & 1 & 1 & \overline{\Yfund} & 1 & -\frac{2}{3} \\
\bar{D}_i & 1 & 1 & \overline{\Yfund} & 1 & \frac{1}{3} \\
\end{array}. \nonumber
\end{equation}

We assume that the first $SU(5)$ gauge group becomes strongly 
interacting around the GUT scale.  This theory then confines with a 
quantum modified constraint given by 
\begin{equation}
\label{QMC}
\det M -B\bar{B}= \Lambda^{10},
\end{equation}
where the mesons are given by $q\bar{q}$ and $q\bar{q}'$, 
both of which acquire a VEV,
while 
the baryons are $q^5$ and $\bar{q}^3 \bar{q}'^2$. In order to ensure
the necessary breaking of the global symmetries to a single
$SU(3)\times SU(2)\times U(1)$ subgroup which is then to be identified 
with the MSSM, one has to make sure that it is the mesons in (\ref{QMC})
which get the expectation values and not the baryons. Note that out of 
only the mesons one can make just a single gauge invariant after the 
$D$-terms of the $SU(5) \times SU(3)\times SU(2)\times U(1)$
are taken into account.  Thus, one expects that once the 
baryon direction is lifted, the vacuum described above is
unique. The baryon directions can be lifted 
by adding a Planck-suppressed superpotential term
\begin{equation}
\frac{1}{M_{\rm Pl}^3} (S q^5 + S' \bar{q}^3 \bar{q}'^2)
\end{equation}
to the action, where $S$ and $S'$ are additional singlets. 
As a consequence, the baryons will get a mass together with the
singlets, and their mass is estimated to be 
$M_{\rm GUT} (M_{\rm GUT}/M_{\rm Pl})^3 
\sim 10^{10}$ GeV. 
Thus we have seen that 
the $N=2$ case can be easily made into a complete model where the
GUT scale emerges dynamically. Of course for $N=2$ one is left with the
question of why the gauge couplings should unify. For this, one has to assume 
that the couplings of $SU(3)\times SU(2) \times U(1)$ are much stronger
than those of the weakly gauged $SU(5)$.

Finally, we remark that this model, just as the continuum versions of these 
models, is ideally set up for gaugino mediation. Gaugino
mediated models in 4D have been constructed in \cite{CEGK,CKSS}. Here
we can simply assume that the SUSY breaking sector only couples to the
endpoint $SU(5)$ site, and is transmitted through the lattice of the $SU(5)$s
to the SM fields at the other end. Since in our case the scale of breaking
to the diagonal gauge group is given by $M_{GUT}$, the gauge mediated
contributions to the scalar masses are only suppressed when the number of
lattice sites is large $N\gtrsim 5$. For this case one obtains a gaugino
mediated spectrum, while for small number of sites one obtains a spectrum that
interpolates between gaugino and gauge mediation.

\section{A Model of Missing Partners}
\label{missing-partner-sec}
\setcounter{equation}{0}
\setcounter{footnote}{0}

\subsection{Setup and doublet-triplet splitting}

While models which
restrict the Higgs to reside at the GUT breaking orbifold fixed-point
give a simple resolution of the doublet-triplet splitting problem
(there are no triplets) such models must sacrifice one of
the most compelling theoretical motivations for GUTs: charge
quantization.  Since in such models
the SM fields must necessarily live at
the same GUT breaking orbifold fixed-point as the Higgs (in order
to get masses from the Higgs VEV) there is no principle which restricts
their $U(1)_Y$ charges to be the observed charges.  For example
different generations  could have different hypercharges (and hence
different electric charges) in this class of models.  If, however,
the Higgs is distributed among all the nodes of the model\footnote{
This is the analog
of being a bulk field in a large $N$ continuum limit (extra dimensional)
model.} then the SM matter fields can reside at any node.  This opens up
many possibilities; most importantly if the SM fields reside at a node
with an unbroken $SU(5)$ gauge group, then the prediction of charge
quantization is restored. However a different resolution of the 
doublet-triplet splitting problem is then required. There are several
ways for this to happen, but they all share the key feature that
the triplets do not have a zero mode\footnote{For an early example
of this idea in a string setting see ref. \cite{Witten}.} 
and thus implement a version of
the missing partner mechanism \cite{missingpartner}.

To be more specific, consider the model shown in Table~\ref{Higgs-model-table} 
(we will discuss the location of the SM generations subsequently).
\begin{table}[t]
\begin{eqnarray*}
\begin{array}{c|ccccccr}
      & SU(5)_1 & SU(5)_2 & \cdots & SU(5)_{N-1} & SU(3) & SU(2) & U(1)_Y  \\
\hline
{\vrule height 15pt depth 5pt width 0pt}
 \bar{P}_{1,\ldots , 5} &
\overline{\Yfund} & 1 & \cdots & 1 & 1 & 1 & 0 \\
Q_1  & \Yfund  & \overline{\Yfund}  & 1       & \cdots & 1 & 1 & 0\\
Q_2  & 1       & \Yfund  & \cdots & 1 & 1 & 1 & 0 \\
        \vdots & \vdots & \vdots & \ddots & \vdots & \vdots & \vdots& \vdots \\
Q_{t,N-1}   & 1 & 1 & \cdots & \Yfund & \overline{\Yfund} & 1 &\frac{1}{3}\\
Q_{d,N-1}   & 1 & 1 & \cdots & \Yfund & 1 & \Yfund & -\frac{1}{2}\\
P_{1,\ldots ,5} & 1 & 1 & \cdots & 1 & \Yfund & 1 & -\frac{1}{3}\\
P'_{1,\ldots ,5} & 1 & 1 & \cdots & 1 & 1 & \Yfund & \frac{1}{2}\\
H_{u,1} & \Yfund & 1 & \cdots & 1 & 1 & 1 & 0 \\
H_{d,1} & \overline{\Yfund} & 1 & \cdots & 1 & 1 & 1 & 0 \\
H_{u,1}^c & \overline{\Yfund}  & 1 & \cdots & 1 & 1 & 1 & 0 \\
H_{d,1}^c & \Yfund & 1 & \cdots & 1 & 1 & 1 & 0 \\
\vdots & \vdots & \vdots & \ddots & \vdots & \vdots & \vdots& \vdots \\
H_{u,N-1} & 1 & 1 & \cdots & \Yfund & 1 & 1 & 0 \\
H_{d,N-1} & 1& 1 & \cdots & \overline{\Yfund}  & 1 & 1 & 0 \\
H_{u,N-1}^c  & 1 & 1 & \cdots & \overline{\Yfund} & 1 & 1 & 0 \\
H_{d,N-1}^c & 1 & 1 & \cdots & \Yfund & 1 & 1  & 0 \\
h_u & 1 & 1 & \cdots & 1 & 1 & \Yfund & \frac{1}{2} \\
h_d & 1 & 1 & \cdots & 1 & 1 & \Yfund & -\frac{1}{2} \\
t_u & 1 & 1 & \cdots & 1 & \Yfund & 1 & -\frac{1}{3} \\
t_d & 1 & 1 & \cdots & 1 & \overline{\Yfund} & 1 & \frac{1}{3} \\
t_u^c & 1 & 1 & \cdots & 1 & \overline{\Yfund} & 1 & \frac{1}{3} \\
t_d^c & 1 & 1 & \cdots & 1 & \Yfund & 1 & -\frac{1}{3} \\
\end{array}. \nonumber
\end{eqnarray*}
\caption{A model of missing partners.}
\label{Higgs-model-table}
\end{table}
There are two ${\cal N}=2$
hypermultiplets of ``bulk'' Higgs fields 
represented by
$H_u$ and $H_d$, transforming at ${\bf 5}$ and $\overline{\bf 5}$
under $SU(5)$, as well as an independent set of fields with conjugate 
quantum numbers $H_u^c$ and $H_d^c$, where $H_u^c$ is in the same
hypermultiplet as $H_u$.
A latticized version of the ``bulk'' Higgs fields have a hopping
term and a mass term in the superpotential:
\begin{equation}
W_{\rm bulk} =   \sum_{i=1}^{N-1} \lambda_i H_{u,i}^c Q_iH_{u,i+1}
               - \sum_{i=1}^{N-1} m_i H_{u,i}^c H_{u,i}
               + u \leftrightarrow d \; .
\end{equation}
As we will discuss later it is useful to have a model with an $R$ symmetry
that forbids mass terms of the form $H_{d}H_{u}$ in order to suppress
proton decay \cite{HN}.  
(This has the added benefit that it forces the 
$\mu$ term to be related to SUSY breaking.) 
For example Hall and Nomura \cite{HN} use an $R$ symmetry with charge $0$ for 
$H_{d}$, $H_{u}$ and charge $2$ for $H_{d}^c$, $H_{u}^c$ 
which accomplishes this.
With generic mass terms and a VEV for $Q$ the Higgs
fields mix and produce four towers of states starting at some non-zero masses.
If we latticize a massless bulk hypermultiplet then the values of $m_i$
are required to be 
proportional to the VEV of $Q$ such that in the large $N$ limit
we recover just the correct kinetic terms.
For a purely 4D model we however do not want to rely on the Lorentz
invariance of the higher dimensional theory, and thus
the bulk mass terms cannot be assumed to be equal.
In the 5D models of SS GUT breaking it is exactly the 5D Lorentz invariance
which enforces the structure of the mass matrices to be such that the
doublets have a zero mode, while the color triplets do not. Since we 
do not want to rely on this symmetry, we have to slightly deviate 
from the strictly deconstructed version of the model in order to avoid fine 
tuning from the 4D point of view. The way we will ensure the existence
of the doublet zero modes is by making sure that the mass matrix for
the doublets is not of maximal rank, by leaving out one of the doublet fields
from each hypermultiplet
in the last (SM) gauge group, while still keeping all the triplet fields.
This way for generic values of mass terms and couplings the triplets 
will all be massive, 
however since
some of the corresponding 
doublets are missing on the fixed-point, the corresponding mass matrix
will not have maximal rank. This is an 
implementation of the  missing partner mechanism 
\cite{missingpartner,Yanagida}, since some doublets are missing and can not
pair up to get a mass term. This way we are resolving the doublet-triplet
splitting problem without fine tuning, however we should emphasize that
this is achieved by a modification of the deconstructed model.
In particular, we are considering the case where the doublets
$h_{d}^c$, $h_{u}^c$ are
missing on the fixed point and we have the superpotential
\begin{equation}
W_{\rm fp}=- m_N
t_{u}^c t_{u}+u \leftrightarrow d ~.
\end{equation}

Let us consider the mass terms for the Higgs fermions
\begin{eqnarray}
\mathcal{L} & \supset& -\sfrac{1}{2}
(\psi_{h,i\alpha}\left| \, \psi^c_{h,i\alpha} )\right. \, 
\delta_{\alpha\alpha'}\,
\left( \begin{array}{r|r}
 & \Xi_h^T \\
\hline
\Xi_h &
\end{array}
\right)_{ij}\,
\left( \begin{array}{r}
\psi_{h,j\alpha'}\\ \hline \psi^c_{h,j\alpha'} \end{array} \right)
\nonumber \\
&&-\sfrac{1}{2}
(\psi_{t,i\alpha}\left| \, \psi^c_{t,i\alpha} )\right. \, 
\delta_{\alpha\alpha'}\,
\left( \begin{array}{r|r}
 & \Xi_t^T \\
\hline
\Xi_t &
\end{array}
\right)_{ij}\,
\left( \begin{array}{r}
\psi_{t,j\alpha'}\\ \hline \psi^c_{t,j\alpha'} \end{array} \right)
+ {\rm h.c.}
\end{eqnarray}
with the $(N-1)\times N$ doublet matrix
\begin{equation}
\Xi_h=
\left( \begin{array}{rrrr}
-m_1 &  \lambda_1 v&\\
&\ddots&\ddots&\\
&&-m_{N-1}&  \lambda_{N-1} v\\
\end{array} \right)
\end{equation}
and the $N \times N$ triplet matrix
\begin{equation}
\Xi_t=
\left( \begin{array}{rrrc}
-m_1 &  \lambda_1 v&\\
&\ddots&\ddots&\\
&&\ddots&  \lambda_{N-1} v\\
&&&-m_N\\
\end{array} \right) \; .
\end{equation}
Unless the $m_i$ and $\lambda_i$ are tuned to special values there 
are no massless triplet fermions
(and by supersymmetry no massless triplet scalars).
For example,
the mass squared matrix of the triplet fermions is
$\Xi_t^T \Xi_t$, which can be easily diagonalized for $m_i=\lambda v$
with $\lambda_1 = \ldots = \lambda_{N-1} = \lambda = \sqrt{2} g$.
Then the (mass)$^2$ spectrum is given by
\begin{equation}
m_k^2= 4 a^{-2} \sin^2\left(  \frac{(2 k+1)\pi}{4 N+2}\right),
\end{equation}
for $k=0\ldots N-1$.
In the models we are considering here there is no reason for such
a special fine tuning so we expect no triplet zero-modes.
However the mass squared matrix for the doublet fermions
$\Xi_h^T \Xi_h$ is an
$N \times N$ matrix of rank $N-1$ so there must be a zero-mode fermion
(and scalar).  Their (mass)$^2$ spectrum is given by
\begin{equation}
m_{h_u,h_d}^2 = 0 \qquad ; \qquad 
m_{h_u,h_d,h_u^c,h_d^c}^2 = 4 a^{-2} \sin^2 \frac{m \pi}{2 N}
\end{equation}
where $m=1 \ldots N-1$.

The additional Higgs doublets and triplets modify the
evolution of the gauge couplings up to $2 a^{-1}$.
The low energy couplings are now
\begin{eqnarray}
\alpha_i^{-1}(M_Z) &=& \alpha_i^{-1}(2 a^{-1}) + \frac{1}{2\pi} 
    \bigg[   b_i \ln \frac{2 a^{-1}}{M_Z}
           + (c_i + d_i + e_i + f_i) (N - 1) \ln 2 \nonumber \\ 
& &{} \qquad\qquad\qquad\quad + f_i \, \ln 2 - \frac{c_i + e_i}{2} \ln N
    \bigg] \; .
\end{eqnarray} 
where $e_i = (6/5,2,0)$ and $f_i = (4/5,0,2)$ correspond 
to the beta function coefficients of two sets of up-type and 
down-type Higgs doublets and triplets, respectively.
The same analysis shown in Sec.~\ref{GCU-sec} can be repeated
with the above evolution equation.  Here, we simply
state the results.  The inverse lattice spacing is 
related to $M_{\rm GUT}$ via
\begin{equation}
2 a^{-1} = 2^{-1/7} N^{2/7} M_{\rm GUT}
\end{equation}
again using $\alpha_1(2 a^{-1}) = \alpha_2(2 a^{-1})$.
We note that the $N^{2/7}$ behavior is {\em identical}\/
to the continuum result found in \cite{HN}, for large $N$.
The deviation of $\alpha_3$ from $\alpha_{\rm GUT}$ is
\begin{equation}
\frac{\alpha_{\rm GUT} - \alpha_3}{\alpha_{\rm GUT}} = 
-\frac{3}{7 \pi} \alpha_{\rm GUT} \ln 8 N \; .
\end{equation}
Unlike in Sec.~\ref{GCU-sec}, here we find the correction is always 
negative, albeit at level of just a few percent.

\subsection{Inclusion of SM matter}

Having dispensed with the doublet-triplet splitting problem we can
now consider the viability of different options for locating the SM
quarks and leptons. As in \cite{HN}, the R-symmetry has to be 
extended to the matter fields such that each SM matter field has
R-charge 1. This is necessary to forbid proton decay from dimension
5 operators.
As in the previous section we can place all the quarks and leptons
on the GUT breaking orbifold fixed-point at the expense of losing
the prediction of charge quantization. It is interesting to note
that Yukawa coupling unification works best for the third generation
(for $\tan\beta \equiv \langle H_u \rangle/\langle H_d \rangle$ large), 
which suggests the third generation is further
from the GUT breaking orbifold fixed-point than the other generations.
It is also interesting to note that intergenerational Yukawa couplings
can be suppressed for generations residing at different nodes, since
gauge invariance will require a factor of the bifundamental $Q$ for
each link that separates the two generations. If such operators
are generated by physics at a scale $M_{\rm flavor}$ within an order of 
magnitude of the GUT
scale, then this leads to realistic predictions for the 
CKM matrix.  (c.f.\ Split-fermion models where a localized Higgs 
wavefunction \cite{Fermi2,splitfermions} generates a hierarchy of Yukawa
couplings.  For our case we have taken the Higgs wavefunction to be
uniform.  The explanation of the hierarchy of the diagonal Yukawa couplings
would require a varying Higgs wave-function along the lattice that is
also possible to implement here with appropriate choices of the coupling 
and mass parameters of the model.)
Recall from the Wolfenstein parameterization of the CKM matrix:
\beq
V_{us} \sim V_{cd} &=& {\cal O}( \lambda)\nonumber\\
V_{cb} \sim V_{ts} &=& {\cal O}( \lambda^2)~.\nonumber
\eeq
This structure is simply reproduced if one link separates the
first and second generation, two links separate the second and
third generations, and $\lambda \sim \langle Q \rangle/M_{\rm flavor}$.  
Provided that the second and
third generation both reside
on the same side of the node with the first generation then this further
implies that the first and third generations are separated by
three links and we naturally
predict the correct order of magnitude for the remaining CKM elements:
\beq
V_{ub}&=& {\cal O}( \lambda^3)\nonumber\\
V_{td}&=& {\cal O}( \lambda^3)~.\nonumber
\eeq

Another alternative for introducing the SM matter in these models is to 
evenly distribute them among the $SU(5)$ gauge groups, corresponding to
having the SM matter in the bulk in the higher dimensional language. In this
case, the matter content has to be doubled, just like we saw for the Higgs 
fields. This is necessary in order to ensure the presence of a massless 
zero mode without fine-tuning. One can obtain this using the method of 
missing partners, where one set of SM fields is omitted at the last site, and 
thus the mass matrix will not be of maximal rank, and a zero mode for the
SM fields is necessarily present. We will not consider this possibility 
any further in this paper.

\subsection{Suppression of proton decay}
The amount of suppression of proton decay crucially depends
on which generations are charged under $SU(5)$.
When all three generations reside at the GUT breaking orbifold 
fixed-point (Fig.~\ref{N-cite-fig}) and there are no dangerous
Higgs triplets, as in the Weiner model \cite{Weiner}, 
proton decay is completely absent.  This is because the
heavy $SU(5)$ gauge bosons ($X, Y$)
do not couple to quarks and leptons.
In models where only the first generation resides at the 
GUT breaking orbifold fixed-point (Fig.~\ref{AB-model-fig}, Model A), 
proton decay will proceed
due to the mixing of the first generation with the other
generations that do couple to heavy $SU(5)$ gauge bosons.
In these models there will be an additional suppression
of these proton decay amplitudes by CKM factors relative
to standard heavy gauge boson processes.  However the leading
contribution to proton decay in the minimal $SU(5)$ SUSY
GUT came from the Higgs triplet fermions, even assuming that they could
be somewhat heavier than the heavy gauge bosons. But in
these models there is nothing that requires the Higgs triplet
to have a direct coupling to the first generation quarks,
so these models are not challenged by the current bounds 
on the proton lifetime. Alternatively if the second and third
generations reside on the GUT breaking orbifold fixed-point
while the first generation resides elsewhere, then proton
decay can still be suppressed since the dangerous operator involves
at least two different generations.  If the triplet couplings to
the second and third generations vanishes, there is no contribution
to proton decay.

As in the Hall and Nomura model, if the Higgs triplet fermion mass is between
$H_a$ and $H_a^c$, but only $H_a$ couples to quarks and leptons 
due to the global R-symmetry, then there
is no dimension 5 operator generated.  
Since the mass of the lightest $X,Y$ gauge bosons is lower
than the usual MSSM GUT scale, we must check, model by model,
that proton decay
mediated by dimension-6 operators does not violate current 
experimental bounds.

\subsection{Two Realistic Models}

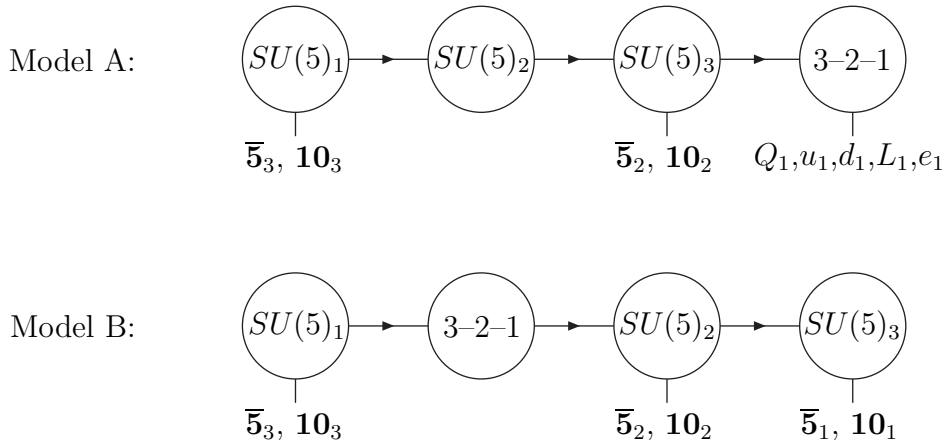
\begin{figure}[!ht]
\begin{picture}(400,60)(0,0)
  \Text(  60, 40 )[r]{Model A:}
  \CArc( 120, 40 )( 20, 0, 360 )
  \Text( 121, 40 )[c]{$SU(5)_1$}
  \ArrowLine( 140, 40 )( 170, 40 )
  \Line( 120, 20 )( 120, 11 )
  \Text( 120,  3 )[c]{${\bf \overline{5}}_3$, ${\bf 10}_3$}
  \CArc( 190, 40 )( 20, 0, 360 )
  \Text( 191, 40 )[c]{$SU(5)_2$}
  \ArrowLine( 210, 40 )( 240, 40 )
  \CArc( 260, 40 )( 20, 0, 360 )
  \Text( 261, 40 )[c]{$SU(5)_3$}
  \ArrowLine( 280, 40 )( 310, 40 )
  \Line( 260, 20 )( 260, 11 )
  \Text( 260,  3 )[c]{${\bf \overline{5}}_2$, ${\bf 10}_2$}
  \CArc( 330, 40 )( 20, 0, 360 )
  \Text( 331, 40 )[c]{3--2--1}
  \Line( 330, 20 )( 330, 11 )
  \Text( 330,  3 )[c]{$Q_1$,$u_1$,$d_1$,$L_1$,$e_1$}
\end{picture}
\begin{picture}(400,100)(0,0)
  \Text(  60, 40 )[r]{Model B:}
  \CArc( 120, 40 )( 20, 0, 360 )
  \Text( 121, 40 )[c]{$SU(5)_1$}
  \ArrowLine( 140, 40 )( 170, 40 )
  \Line( 120, 20 )( 120, 11 )
  \Text( 120,  3 )[c]{${\bf \overline{5}}_3$, ${\bf 10}_3$}
  \CArc( 190, 40 )( 20, 0, 360 )
  \Text( 191, 40 )[c]{3--2--1}
  \ArrowLine( 210, 40 )( 240, 40 )
  \CArc( 260, 40 )( 20, 0, 360 )
  \Text( 261, 40 )[c]{$SU(5)_2$}
  \ArrowLine( 280, 40 )( 310, 40 )
  \Line( 260, 20 )( 260, 11 )
  \Text( 260,  3 )[c]{${\bf \overline{5}}_2$, ${\bf 10}_2$}
  \CArc( 330, 40 )( 20, 0, 360 )
  \Text( 331, 40 )[c]{$SU(5)_3$}
  \Line( 330, 20 )( 330, 11 )
  \Text( 330,  3 )[c]{${\bf \overline{5}}_1$, ${\bf 10}_1$}
\end{picture}
\caption{Diagrammatic illustration of the two models.}
\label{AB-model-fig}
\end{figure}

In this section we will discuss two simple realistic models.
The field content of bifundamentals ($Q$), spectators ($P$),
and Higgses ($H$) is along the lines discussed in the previous
sections.  The distributions of quarks and leptons (as well
as squarks and sleptons) is depicted schematically in 
Fig.~\ref{AB-model-fig}.

Both models implement the missing doublet scenario described earlier.
In model A the first generation resides at the GUT breaking orbifold
fixed-point, while the second generation resides on the adjacent 
GUT symmetric site, and the third generation resides on the site 
two links away from the second generation.
Note
that the spacing of the generations qualitatively reproduces the CKM 
matrix as discussed above. Also the third generation is the furthest
away from the source of GUT breaking so this model can naturally
incorporate  a more accurate Yukawa unification
in the third generation, while
there is no reason to expect Yukawa unification in the first generation.
As discussed above an $R$ symmetry prevents the coupling of
SM fields to triplets necessary to generate dimension 5 proton decay 
operators, so the leading contribution to proton decay is through
$X$, $Y$ gauge boson exchange.  Since the first generation does not
directly couple to $SU(5)$ gauge bosons, these couplings are
only generated through mixing with the other generations.  Thus there
is an additional CKM suppression (changing flavor to the second generation)
of these decays relative to
usual GUTs, so this model is phenomenologically viable. This model also has
the signature that proton decays with electrons in the final state are
highly suppressed (they arise only through lepton flavor violating effects).
Experimentally the limit on proton decays to second generation particles is 
considerably weaker than that for first generation decays.

 The main theoretical blemish of model A is that charge quantization
is not guaranteed.  However we can simply overcome this problem by
carrying over the virtues of this model to a similar model which
cannot be obtained by latticizing a 5D orbifold.
Model B is just such a model:  in this model all three generations
reside at GUT symmetric sites, while the GUT breaking site sits in
the middle.  Thus charge quantization is a prediction of model B.
Furthermore, the missing partner mechanism can be implemented in
a straightforward manner (following the discussion above)
and the spacing between generations
is maintained so the CKM structure is maintained.  Dimension
6 proton decay operators no longer have an additional CKM suppression,
but the model is not in conflict with the experimental limit
since the number of lattice sites, $N=4$, is small.
Since in Model B proton decay can proceed
directly to first generation particles, the bound
on proton decay via $\tau_{p\rightarrow e^+ \pi^0} > 1.6 \times 10^{33}$ 
years is the most stringent.  In particular, the decay through a 
heavy $X$ gauge boson gives
\begin{equation}
\tau_{p\rightarrow e^+ \pi^0}
 \sim 10^{33}\, {\rm years}\,\left({{M_X}\over{5 \times
10^{15} \,{\rm GeV}}}\right)^4
\end{equation}
and so these operators
are safe so long as $M_X \gtrsim 5 \times 10^{15}$ GeV, which
requires $N \lesssim 4$.

\section{Conclusions}
\setcounter{equation}{0}
\setcounter{footnote}{0}

We have considered 4D supersymmetric models that
predict the unification of couplings at a high scale. These models were
inspired by the deconstruction of 5D $SU(5)$ gauge theories, where the
GUT group is broken by the Scherk-Schwarz mechanism 
on an $S^1/Z_2\times Z'_2$ orbifold. The 4D models consist
of a chain of $SU(5)$ gauge groups, with one site being replaced by
the SM gauge group, and bifundamental fields that break the full gauge 
symmetry to the diagonal SM group. In the simplest model, the SM matter 
is included at the last site where the gauge symmetry is reduced, and thus
the doublet-triplet splitting problem is automatically resolved, while
proton decay is exactly zero. However, this model does not explain the
quantum numbers of the SM fields. A slightly modified version has the
SM fields at $SU(5)$ sites, and the Higgs fields are distributed among all
the sites. These models do explain charge quantization, and the 
doublet-triplet splitting can be resolved by a very simple
version of the missing partner
mechanism. Proton decay is now reintroduced, but only through the dimension
6 operators, which can be sufficiently suppressed as long as the number
of sites is not too large. By putting the SM fields at different sites,
a realistic hierarchy of the CKM matrix elements can be obtained.

\section*{Acknowledgements}

We thank Nima Arkani-Hamed, Josh Erlich, Aaron Grant, Christophe Grojean, 
Yasunori Nomura, and Neal Weiner
for many useful discussions.
C.C. and G.D.K. also thank the Aspen Center for Physics where part of this work
was completed.
C.C. is an Oppenheimer fellow at the Los Alamos National Laboratory.
C.C. is supported by
the U.S. Department of Energy under contract W-7405-ENG-36. 
G.D.K. 
is supported in part by the U.S. Department of Energy 
under contract DE-FG02-95-ER40896.
J.T. is supported in part by the NSF under grant PHY-98-02709.


\end{document}